# Evolution of coherence between the Jaynes – Cummings Eigenstates in the presence of repeated occupancy measurements: The Case of Resonant Coupling

**Vikram S. Athalye**

Indian Institute of Science Education and Research, 900, NCL Innovation Park, Dr. Homi Bhabha Road, Pune – 411 008, India

E-mail: vs.athalye@iiserpune.ac.in

**Abstract**

A two – level system resonantly coupled to a single mode cavity field and subject to ground state occupancy measurement – like interaction is considered. For this situation, the solution to the Lindblad master equation for the density matrix is obtained with respect to the Jaynes – Cummings eigenstate representation. It is seen that, the rate of decoherence of the superposition of the Jaynes – Cummings eigenstates increases monotonically with the increase in the occupancy measurement coupling up to a certain critical value (already reported in the literature in a different context), whereas the rate decreases with the increase in the measurement coupling beyond the same critical value.

## 1. A two – level system in a single mode cavity

The 'real time observation' of the dynamics of a single quantum system and the study of the evolution of quantum coherence between its relevant states in the presence of external perturbations require its isolation and confinement for a long time [1]. The isolation and confinement, followed by the ability to manipulate the quantum states of an individual system and to perform repeated measurements on it, are necessary for exploring the connection between the quantum and the classical physics [2], for understanding the counter – intuitive notion of entanglement [3], and for implementing complex algorithms in information science according to quantum logic [4]. In recent years, the field of experimental cavity QED has demonstrated ever improving capabilities, especially in the strong coupling regime, to accomplish these tasks [5-7].

The quantum dynamics of a two – level system, with $|g'\rangle$ and $|e'\rangle$ as the ground and excited states, trapped in a single mode cavity, is given by the Jaynes – Cummings Hamiltonian [8]

$$\hat{H} = \frac{1}{2}\hbar\omega_S\hat{\sigma}_z \otimes \hat{I}_F + \hat{I}_S \otimes \hbar\omega_F\left(\hat{a}^\dagger\hat{a} + \frac{1}{2}\right) + \hbar g\left(\hat{a}^\dagger\hat{\sigma} + \hat{a}\hat{\sigma}^\dagger\right) \tag{1}$$

Here, $\hat{\sigma}_z, \hat{\sigma}^\dagger$ and $\hat{\sigma}$, belonging to the set of Pauli operators, correspond to inversion, raising and lowering of the state of the two – level system, $\hat{a}^\dagger$ and $\hat{a}$ are the single mode creation and annihilation operators acting on field kets, $\omega_S$ and $\omega_F$ are the resonant frequencies for the two – level system and the cavity field, and g is the effective coupling constant for the interaction between them. For the case of resonant coupling $(\omega_S = \omega_F)$, the eigenstates of Hamiltonian (1) are the Jaynes – Cummings eigenstates (JCE) corresponding to symmetric and antisymmetric combinations of the states $|g', n\rangle$ and $|e', n-1\rangle$, given by

$$|\pm\rangle_n = \frac{1}{\sqrt{2}}\left[\left|g', n\right\rangle \pm \left|e', n-1\right\rangle\right] \tag{2}$$

where, there are $n$ quanta of excitation in the cavity QED system. The present work assumes a cavity QED situation in the strong coupling regime. In this regime the coherent interaction between the system and the cavity field dominates the usual dissipative processes such as the escape of photons through the cavity mirrors and the incoherent decay of the excited state of the two – level system due to its coupling to the free – space electromagnetic background [7]. Consequently, the cavity field along with the two – level system behaves as a *combined* quantum system with the JCE given by equation (2) as its *relevant* stationary nondegenerate energy eigenstates with eigenvalues

$$E_{\pm} = n\hbar\omega \pm \hbar\sqrt{n}g \ \left(\text{with } \omega = \omega_S = \omega_F\right) \tag{3}$$

Recently reported investigation of the energy level structure of such a quantum system, by averaging over $\sim 10^3$ atoms [9] and the observation of the vacuum Rabi spectrum [corresponding to the normal mode splitting with $n = 1$ in equations (2) and (3)] for *one* trapped atom by implementing a novel scheme for cooling its axial motion [10] confirm its experimental accessibility.

In this analysis, the effect of repeated ground state occupancy measurements, to be performed at the trapped two – level system, on the coherence between these JCE is investigated by using the Lindblad form of the master equation for the density operator of the combined system [11-12]. As proposed by Cook for the case of a single trapped ion to demonstrate the quantum Zeno effect [13] and realized later with an ensemble of about 5000 $Be^+$ ions confined in a Penning trap by Itano *et al* in the presence of classical Rabi oscillations [14], such measurements could be performed by using appropriate number of optical pulses of appropriately chosen duration leading to the spontaneous recurring dipole transitions $|e''\rangle \rightarrow |g'\rangle$ from a higher excited state $|e''\rangle$, with the transition $|e''\rangle \rightarrow |e'\rangle$ forbidden. The corresponding theoretical analysis in the conventional stationary state representation (i.e. in the eigenstate representation of the unperturbed Hamiltonian for the two – level system) has been reported in the literature [15]. However, as mentioned above, in the strong (resonant) coupling regime of cavity QED the naturally existing stationary states of the combined system are the JCE expressed by equation (2) or more generally the so – called dressed states (in the presence of non – resonant coupling) [16]. It can therefore be interesting to study the evolution of coherence between the states that represent the *entanglement between the two – level system and the cavity field*, in the presence of repeated measurements performed on the trapped two – level system. This is done in the next section for the case of repeated occupancy measurements. In section 3, a review of and discussion on the result obtained in the next section are presented.

## 2. The master equation and its solution

For the situation considered in this letter, the Lindblad form of the master equation is written as:

$$\frac{d\hat{\rho}}{dt} = -\frac{i}{\hbar}\left[\hat{H}, \hat{\rho}\right] - \frac{\kappa}{2}\left[\hat{A}, \left[\hat{A}, \hat{\rho}\right]\right] \tag{4}$$

Here, $\hat{H}$ is given by equation (1), $\kappa$ is the measurement coupling taken to be proportional to the rate of spontaneous transition $|e''\rangle \rightarrow |g'\rangle$ with $\hat{A}$ as the Lindblad operator related with the occupancy measurements on $|g', n\rangle$. By defining $\hat{A}|g', n\rangle = |g', n\rangle$ and $\hat{A}|e', n-1\rangle = 0$ and by using equation (2), it can be easily checked that its matrix elements in the Jaynes − Cummings eigenbasis are

$$A_{++} = A_{+-} = A_{-+} = A_{--} = \frac{1}{2} \tag{5}$$

Using these matrix elements in equation (4), we get the following system of differential equations:

$$\begin{aligned}
\dot{\rho}_{++} &= -\frac{\kappa}{4}\left[\rho_{++} - \rho_{--}\right] \\
\dot{\rho}_{--} &= +\frac{\kappa}{4}\left[\rho_{++} - \rho_{--}\right] \\
\dot{\rho}_{+-} &= \left[-iR - \frac{\kappa}{4}\right]\rho_{+-} + \frac{\kappa}{4}\rho_{-+} \\
\dot{\rho}_{-+} &= \left[+iR - \frac{\kappa}{4}\right]\rho_{-+} + \frac{\kappa}{4}\rho_{+-}
\end{aligned} \tag{6}$$

Here, $R = 2g\sqrt{n}$ is the $n$ − photon Rabi frequency. The solutions to this system subject to the initial condition

$\rho_{++}(0) = \rho_{--}(0) = \rho_{+-}(0) = \rho_{-+}(0) = \frac{1}{2}$ and to the requirement of trace preservation, $\mathrm{Tr}[\hat{\rho}] = 1$ are given by

$$\rho_{++}(t) = \rho_{--}(t) = \frac{1}{2} \tag{7a}$$

$$\rho_{+-}(t) = \frac{\kappa + \sqrt{\kappa^2 - (4R)^2}}{4\sqrt{\kappa^2 - (4R)^2}} \exp\left\{-\frac{\kappa}{4} + \frac{\sqrt{\kappa^2 - (4R)^2}}{4}\right\}t$$

$$-\frac{\kappa - \sqrt{\kappa^2 - (4R)^2}}{4\sqrt{\kappa^2 - (4R)^2}} \exp\left\{-\frac{\kappa}{4} - \frac{\sqrt{\kappa^2 - (4R)^2}}{4}\right\}t$$

$$-\frac{iR}{\sqrt{\kappa^2 - (4R)^2}} \times \qquad\qquad (7b)$$

$$\left(\exp\left\{-\frac{\kappa}{4} + \frac{\sqrt{\kappa^2 - (4R)^2}}{4}\right\}t - \exp\left\{-\frac{\kappa}{4} - \frac{\sqrt{\kappa^2 - (4R)^2}}{4}\right\}t\right)$$

and

$$\rho_{-+}(t) = \rho_{+-}(t)^* \qquad\qquad (7c)$$

In the absence of the occupancy measurements, i.e. for $\kappa = 0$, we get

$$\rho_{+-}(t) = \frac{1}{2}\exp(-iRt) = \rho_{-+}(t)^* \qquad\qquad (8a)$$

This oscillatory behavior of coherence between the JCE corresponds to time – varying entanglement between the two – level system and the cavity field. On the other hand, in the limit $\kappa \to \infty$ one finds that

$$\rho_{+-}(t) \to \frac{1}{2} \qquad\qquad (8b)$$

Clearly, in both these limits the expressions for coherence between the JCE as given by equations (8a) and (8b) are as expected. However, it is seen from equation (7b) that for $\kappa \in (0, 4R)$, an increase in $\kappa$ leads to an *increase* in the rate of decoherence of the superposition of JCE. Whereas, an increase in $\kappa$ in the domain $(4R, \infty)$ leads to a *decrease* in decoherence rate. These two different ways in which the evolution of the coherence of JCE takes place in the presence of the occupancy measurement – like interaction are seen to be separated by a critical value of the occupancy measurement coupling, namely $\kappa_{crit} = 4R$ [15]. Below we show different plots of the evolution of coherence between the JCE for different values of the occupancy measurement coupling (below and above its critical value) expressed in terms of the Rabi frequency, with $R = 100$ MHz, over a timescale of 0.1 μs:

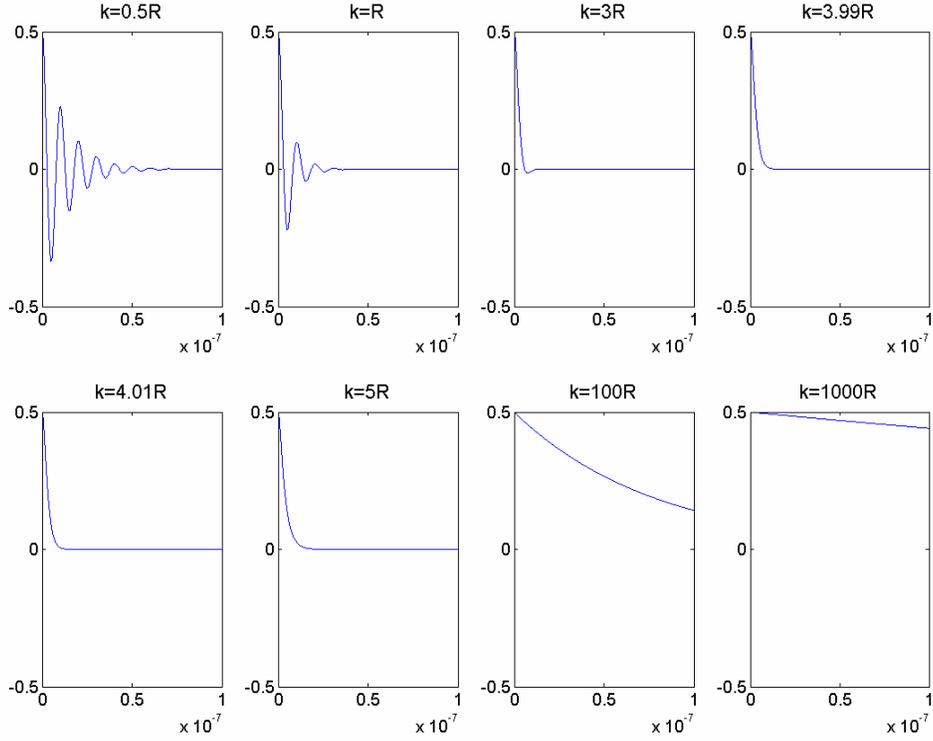

## 3. Discussion

A measurement – like interaction, depending on its nature, involves a certain choice of basis in which it has a *diagonal* representation and causes the local destruction of phase relations between the quantum states belonging to *that* basis [17-18]. The rate of decoherence of superposition of the quantum states belonging to the chosen basis always *increase* monotonically with the increase in the measurement coupling or the frequency of measurements. Clearly, the occupancy measurement – like interaction has a diagonal representation in the eigenstates of the unperturbed part of the Hamiltonian given by equation (1). Also, the rate of decoherence of superposition of these eigenstates does follow a monotonic increase with the increase in the occupancy measurement coupling [15]. However, in this investigation it is shown that, in the presence of Rabi oscillations, the same interaction leads to decoherence of superposition of quantum states (JCE) belonging to that basis in which it *does not* have a diagonal representation [see equation (5)]. It is also noticed that the rate of decoherence does not follow a monotonic increase for all values of the measurement coupling $\kappa$ within the range $(0, \infty)$, but rather increases monotonically up to a critical measurement coupling ($\kappa_{crit} = 4R$) and starts decreasing beyond that. Thus, equation (7b) represents the main result of the analysis, which describes a clearly *counterintuitive* evolution of coherence between the JCE under the influence of repeated occupancy measurements. It predicts that, such measurements would destroy the coherence between the JCE over a time scale $(T)_<^{dec} \sim \dfrac{4}{\kappa}$ for $\kappa < \kappa_{crit}$ and over a time scale $(T)_>^{dec} \sim \dfrac{4}{\sqrt{\kappa^2 - (4R)^2} - \kappa}$ for $\kappa > \kappa_{crit}$. Such an analysis could also be relevant in the context of quantum feedback control [19].

**Acknowledgements**

I wish to thank Arvind Kumar of HBCSE, Tata Institute of Fundamental Research, Mumbai, India and T. S. Mahesh of IISER, Pune, India for many helpful discussions.

_______________________________